\documentstyle[11pt]{article}

\newcommand{\be}{\begin{equation}}
\newcommand{\ee}{\end{equation}}
\newcommand{\bea}{\begin{array}}
\newcommand{\ea}{\end{array}}
\newcommand{\beqa}{\begin{eqnarray}}
\newcommand{\eeqa}{\end{eqnarray}}
\newcommand{\bean}{\begin{eqnarray*}}
\newcommand{\eean}{\end{eqnarray*}}

\def\up#1{\leavevmode \raise.16ex\hbox{#1}}

\def\sqr#1#2{{\vcenter{\vbox{\hrule height.#2pt
        \hbox{\vrule width.#2pt height#1pt \kern#1pt
          \vrule width.#2pt}
        \hrule height.#2pt}}}}

\newcommand{\gapproxeq}{\lower .7ex\hbox{$\;\stackrel{\textstyle >}{\sim}\;$}}
\newcommand{\lapproxeq}{\lower .7ex\hbox{$\;\stackrel{\textstyle <}{\sim}\;$}}

\def\thebibliography#1{{\bf REFERENCES\markboth
 {REFERENCES}{REFERENCES}}\list
 {[\arabic{enumi}]}{\settowidth\labelwidth{[#1]}\leftmargin\labelwidth
 \advance\leftmargin\labelsep
 \usecounter{enumi}}
 \def\newblock{\hskip .11em plus .33em minus -.07em}
 \sloppy
 \sfcode`\.=1000\relax}

\begin{document}
\begin{flushright}
DSF 16-2002  \\ quant-pg/0208024\\
\end{flushright}
\vspace{1cm}

\begin{center}
{\bf \large{COMPLEMENTARITY, MEASUREMENT AND INFORMATION

IN INTERFERENCE EXPERIMENTS }}\\
\vspace{1cm}
G. Bimonte and R. Musto
\end{center}
\vspace{1cm}

\begin{center}
{\it   Dipartimento di Scienze Fisiche, Universit\`{a} di Napoli, Federico
II\\Complesso Universitario MSA, via Cintia, I-80126, Napoli, Italy;
\\ INFN, Sezione di Napoli, Napoli, ITALY.\\
\small e-mail: \tt bimonte,musto@napoli.infn.it } \\
\end{center}

\begin{abstract}
Different criteria (Shannon's entropy, Bayes' average cost, D\"{u}rr's
normalized rms spread) have been introduced to measure the "which-way"
information present in interference experiments where, due to
non-orthogonality of the detector states,  the path determination is
incomplete. For each of these criteria, we determine the optimal
measurement to be carried on the detectors, in order to read out the
maximum which-way information. We show that, while in two-beam experiments,
the optimal measurement is always provided by an observable involving the
detector only, in multibeam experiments, with equally populated beams and
two-state detectors,
 this is the case only for the D\"{u}rr criterion, as the other two
require the introduction of an ancillary quantum system, as part of the
read-out apparatus.

\end{abstract}

\section{Introduction}

The debate on double-slit interference experiments,  with photons or
matter particles, and on the possibility of detecting, as proposed by
Einstein, "which-way" individual particles are taking, helped to shape
the basic concept of complementarity in quantum mechanics. According
to this early discussion, Young interference experiments were showing
the wave nature of both radiation and matter and  any attempt to
exhibit their, complementary,
 particle nature, by detecting which path each an individual quantum was
travelling, was regarded as implying a disturbance capable of destroying
the interference pattern. \footnote{For an analysis of the Bohr-Einstein
dialogue and reprints of the relevant papers  see \cite{wheeler}} It was,
however, much later noticed that "in Einstein's version of the double-slit
experiment, one can retain a surprisingly strong interference pattern by
not insisting on a 100\% reliable determination of the slit through which
each photon passes"\cite{zurek}.

More recently this problem has been thoroughly investigated both from a
theoretical and an experimental point of view, by proposing {\it
gedanken}-experiments or actually performing them, in which the quantum
unitary evolution of both the system and the detector is completely under
control. In many cases  care is taken of having the detectors acting on
internal degrees of freedom, so that they do not disturb directly the
centre of mass motion.

 As it is well known  the partial loss of contrast of the interference fringes,
 their modification or total disappearance, find a complete
 quantum mechanical description  in terms
 of the entanglement between the  interfering particles and
 the detectors. To be more precise, the unitary
evolution describing  the interaction of the system with the detectors
leads to the entangled state
\begin{equation}\label{Psit}
  |\Psi(t)> = |\psi_1(t)>\otimes\,|\chi_1> + |\psi_2(t)>\otimes
  \,|\chi_2>\;,\label{preme}
\end{equation}
where  $|\psi_i>$ $i=1,2$, denote the states of the beams going through
slits $1$ and $2$, respectively, while $|\chi_i>$, $i=1,2$ are the
(normalized) detector final states, and $t$ is any time after the system
has left the detection region. The structure of the interference fringes
may be read off the probability density on the screen:
\begin{equation}
   |\!<\!x|\Psi(t_1)\!>\!|^2 =|\!<\!x|\psi_1(t_1)>\!|^2 +\,
  |\!<\!x|\psi_2(t_1)>\!|^2 +2 Re\{<\!\psi_1(t_1)|x\!><\!x|\psi_2(t_1)\!>
<\!\chi_1|\chi_2\!>\}\;.\label{inter}
\end{equation}
Depending on the value of $<\!\chi_1|\chi_2\!>$ there is a continuum
between the extreme cases of no which-way detection
($|\chi_1\!>=|\chi_2\!>$), where the wave nature is exhibited by
interference fringes with maximum contrast, and perfect which-way detection
($<\!\chi_1|\chi_2\!>=0$), where the interference fringes disappear. For
example, in the experimental realization \cite{chap} of Feynman's {\it
gedanken}-experiment \cite{feyn},
 the states $|\chi_i>$ describe the  scattered
photon needed to detect whether  the atom (rather than the electron, as in
the original discussion) has passed through slit $1$ or $2$ and the
quantity $<\chi_1|\chi_2>$  can be varied by changing the spatial
separation between the interfering paths at the point of scattering. In the
experimental setup proposed in \cite{scully} the which-way detection is
performed by micro-maser cavities inserted on the beams of previously
exited atoms. Atomic decay in one of the cavities provides a which-way
information whose predictability depends on the initial state of the
cavities. However we should point out that the detector needs not be a
separate physical system: the which-way information may indeed be stored in
some internal degrees of freedom of the interfering particles, as it
happens in neutron interference experiments \cite{neutrons}, where the spin
of the neutron in one of the beams  is rotated with respect to the original
common direction. Notice that, in each of these examples, the structure of
the interference fringes, as it is clear from Eq. (\ref{inter}), depends on
the entanglement of the system with the apparatus, from which a "which-way"
information may be eventually recovered by means of an appropriate
measurement, and not on the fact of actually performing it. Eq.
(\ref{preme}) describes only a premeasurement.
 Therefore the actual measurement
relative to the "which-way" information may be arbitrarily delayed. As
Schr\" odinger puts it, in his "general confession" \cite{sch}, motivated
by the appearance of the Einstein, Podolsky, Rosen paper \cite{EPR},
"entanglement of predictions" goes "back to the fact that the two bodies at
some earlier time formed in a true sense {\it one} system, that is were
interacting, and have left behind {\it traces} on each other".

Furthermore, it should be stressed that, apart from the extreme  case in
which $<\chi_1|\chi_2>=0$, no   measurement can provide full information on
the way that an individual quantum has taken.
 One is actually
dealing with a problem in quantum detection theory, that is, in statistical
decision theory. In order to decide  what measurement should be carried out
to extract the best possible which-way information, it is necessary to
spell out a strategy in which an {\it a priori} evaluation criterion is
given.

In the pioneering work of Wootters and Zurek, \cite{zurek}, Shannon's
definition of information entropy \cite{shannon} was taken as a
quantitative  measure of the gain in  "which-way" information obtained by
actually performing a measurement on the detector state. In this framework
evidence was produced that "the more clearly we wish to observe the wave
nature ...the most information we must give up about its particle
properties". Following this suggestion, Englert \cite{englert}, by using a
different criterion for evaluating the available information, was able to
establish, for equally populated beams, a complementarity relationship
between the {\it distinguishability}, that gives a quantitative estimate of
the ways, and the {\it visibility} that measures the quality of the
interference fringes:

\begin{equation}
  {\cal D}^2 +{\cal V}^2 \leq 1\;,\label{D+V}
\end{equation}
with equality sign holding if the detector is prepared in a pure state.
\noindent
As usual $\cal V$ is defined in terms of the maximum  and minimum intensity
of the fringes ($I_M$ and $I_m$), ${\cal V}= (I_M - I_m)/(I_M + I_m)$.
$\cal D$ is simply related to
the optimum average Bayes's cost $\bar{C}_{opt}$, traditionally used in
decision theory, by the relation ${\cal D}=1-2\,\bar{C}_{opt}$
\footnote{In Ref. \cite{englert} the distinguishability is expressed in terms
of the optimum likelihood ${\cal L}_{opt}$ for "guessing the way right".
This optimum likelihood is one minus the optimum average Bayes cost
$\bar{C}_{opt}$}.

New problems arise in going from the case of two beams to a multibeams
interference process. As shown by D\"{u}rr \cite{durr}, the complementarity
relationship [Eq. (\ref{D+V})] still holds when the visibility and  the
distinguishability are taken to be,  the first as the, properly normalized,
deviation of the fringes intensity from its mean value, and the second,
following an alternative notion of entropy introduced in Ref. \cite{bruk},
as the maximum  average rms spread of the a posteriori probabilities for
the different paths (see Sec. 2).

The purpose of this paper is to examine an interesting physical aspect of
the problem, that seems to have been overlooked, so far, and it is the
following: once a specific criterion to measure the which-way information
is chosen, what is the actual measurement that has to be performed on the
detectors, in order to extract the optimum information? The usual attitude
to address this question, is to consider the set ${\cal A}_D$ of all
observables $A$, relative to the detector, and to search, among them, for
the observable that delivers most information. However, it is known from
quantum detection theory \cite{helstrom,peres}, that the amount of
information that can be obtained in this way does not represent, in
general, the absolute maximum. Sometimes, it is  possible to do a better
job by introducing, in addition to the detector, an {\it ancilla}, namely
an auxiliary quantum system, neither interacting with the detector, nor
having any correlation with it. Despite the fact that the detector and the
ancilla are, under all respects, independent systems, it may happen that a
{\it larger} amount of information can be obtained, by measuring an
observable relative to the combined system. In connection with this issue,
we point out that, even if the quantity $\cal D$ appearing in Eq.
(\ref{D+V}) is usually defined in relation with ${\cal A}_D$, the proofs
leading to Eq. (\ref{D+V}), say in Refs. \cite{englert, durr}, remain valid
if one includes the observables for the system formed by the detector and
the ancilla together. It follows that the quantity ${\cal D}$ really refers
to all possible detector+ancilla systems.

Since the need for an ancilla seems to us a source of  undesirable
complication for the read out apparatus, it would be interesting to know
under what circumstances the ancilla is really required. In particular, it
would be interesting to know if  there exist  criteria to measure the
which-way information, such that the optimal measurement turns out to be an
ordinary observable relative to the detector,  and the inclusion of an
ancilla does not lead to any improvement. We show that, in the case of
two-beams interference experiments, with either one of the two proposed
measures of information, the optimal measurement does not involve an
ancilla. On the contrary, in the case of multibeam experiments, it is only
with the criterion introduced in Ref. \cite{durr} that the ancilla is
unnecessary, while it is required for the other two criteria, in general.
It is interesting to notice that the criterion for which ordinary
measurements are good enough is the one that leads to the complementarity
relation given by Eq. (\ref{D+V}). Finally, let us notice that, while
inspired by the problem of complementarity in interference experiments, our
work is a contribution to the difficult problem of optimization in quantum
decision theory.

The paper is organized as follows. In Sec. 2 the quantum detection problem
for non-mutually orthogonal detector states is presented and the notion of
ancilla is introduced. We review a fundamental theorem by Neumark, stating
that measurements involving  an ancilla in the enlarged detector-ancilla
Hilbert space, can be equivalently described by means of {\it positive
operator-valued measures} (POVM) on the detectors's Hilbert space,
generalizing the ordinary {\it projection-valued measures} (PVM), that
describe measurements   not involving the ancilla. We then list the
conditions that must be satisfied by any function, for it to be a good
measure of the amount of information provided by a POVM. The different
choices present in the literature for such a function are considered, and
the resulting optimization problems are studied in Sec. 3, for the case of
two beams, and  in Sec. 4 for multibeam interferometers. Some of the proofs
are postponed to an Appendix. Final remarks and a discussion of
perspectives close the paper.

\section{The quantum decision problem.}
\setcounter{equation}{0}

We consider a $n$-beam interference experiment: a single beam of
identical microscopic systems, like photons, electrons, neutrons,
atoms etc. (generically  referred to as particles), is divided into
$n$ spatially separated beams by some sort of beam-splitter, like a
screen with $n$ slits. The $n$ beams are then recombined on a screen,
and the interference figure is observed. It is assumed that the
intensity of the beam is adjusted so that only one particle at a time
passes trough the interferometer, and that the populations $\zeta_i$
of each of the $n$ beams can be adjusted at will. We imagine now that
a detector, designed to provide which-way information on individual
particles passing through the interferometer, is placed along the
trajectories of the beams. It is assumed that the detector also can be
treated as a quantum system, and that the system-detector interaction
gives rise to some unitary process. The detector will serve as
which-way detector if, once prepared in some fixed state $|\chi_0\!>$,
it is brought by the interaction with the particles into a new state,
that depends on the beam occupied by the particle. In formulae, this
amounts to requiring that, after the interaction, the state of the
particle-detector system is the following entangled state,
generalizing [Eq. (\ref{preme})]:
\be
\sum_{i=1}^n c_i \;|\psi_i\!>\otimes \,|\chi_i\!>\;.\label{nbeam}
\ee
Here, $|\psi_i\!>$ denote the normalized particles wave-functions for the
individual beams, while $|\chi_i\!>$  are $n$ normalized (but not
necessarily orthogonal !) states of the which-way detectors. We {\it
define} the detector's Hilbert space ${\cal H}_D$ as the linear span of the
states $|\chi_i>$:
\be
{\cal H}_D := {\rm span}\{|\chi_i>\;,\;i=1,\dots n\}\;.
\ee
(Of course, it may very well happen that the set  of {\it all} possible
states of the detector, as a physical system, is actually larger than
${\cal H}_D$.) In concrete experiments $|\chi_i\!>$ may in fact be internal
states of the particles themselves, in which case $|\psi_i\!>$ denotes the
space-part of the particles wavefunction. We assume that the amplitudes
$c_i$ are known in advance, such that the weights $\zeta_i=|c_i|^2$ give
the a priori probabilities for a particle to pass through the $i$-th slit.
The state [Eq. (\ref{nbeam})] describes a situation in which there is
complete correlation between the beams and the internal states of the
detector, such that, if the detector is found to be in the state
$|\chi_i\!>$, one can tell with certainty that the particle passed through
the $i$-th slit. Thus the problem of determining the trajectory of the
particle reduces to the following one: after the passage of each particle,
is there a way to decide in which of the $n$ states $|\chi_i\!>$ the
detector was left?  If the states $|\chi_i\!>$ are orthogonal to each
other, the answer is obviously yes. Indeed, if we let ${\cal A}_D$ the set
of all hermitean operators in ${\cal H}_D$,  we can surely find in ${\cal
A}_D$ an observable $A$ , such that:
\be
A \;|\chi_i>= \lambda_i \;|\chi_i>\;,\;\;\;\;\;\lambda_i \neq \lambda_j\;
\;\;\;{\rm for}\;
i \neq j\;.\label{orto}
\ee
If $A$ is measured, and the result $\lambda_i$ is found, one can infer with
certainty that the detector was in the state $|\chi_i>$. If, however, the
states $|\chi_i\!>$ are not orthogonal to each other, for no choices of $A$
one can fulfil Eq. (\ref{orto}): whichever $A$ one picks, there will be at
least one eigenvector of $A$, having a non-zero projection onto more than
one state $|\chi_i\!>$. Therefore, when the corresponding eigenvalue is
obtained as the result of a measurement, no unique detector-state can be
inferred,
 and only probabilistic judgments
can be made. Under such circumstances, the best one can do is to select the
observable that provides  as much information as possible, ${\it on \; the
\;average}$, namely after many repetitions of the experiment.
Of course, this presupposes the choice a definite criterion to measure
the average amount $\bar F{(A)}$ of which-way information delivered by
a certain observable $A$ (the properties of $\bar{F}(A)$, and the
various choices proposed so far for this quantity are discussed later
in this Section). After this choice is made, the {\it
distinguishability} ${\cal D}$ of the trajectories is usually related
to the supremum, $F_{D}$, of $\bar F{(A)}$, over ${\cal A}_D$.

It may now come as a surprise to notice, as pointed out in the
Introduction, that the quantity $F_D$  does not always represent the
absolute maximum information that is actually available. Indeed, it is an
intriguing feature of the quantum detection problem, for non orthogonal
states, that a {\it larger} amount of information on the state of the
detector can be obtained by considering the detector in {\it combination}
with an auxiliary quantum system, called {\it ancilla} \cite{helstrom,
peres}. The ancilla does not interact with the detector, and is prepared in
a fixed known state $ |\phi_{0}>\in {\cal H}_{aux}$, such that the combined
system is in one of the $n$ uncorrelated states $|\chi_i>\otimes
\;|\phi_{0}>$, belonging to the total Hilbert space ${\cal H}_{tot}={\cal H}_D
\otimes {\cal H}_{aux}$. Let now ${\cal A}_{tot}$ the set of all hermitean
operators in ${\cal H}_{tot}$ and  $F_{tot}$ the supremum of ${\bar
F}(A)$ over ${\cal A}_{tot}$
. Surprisingly enough, even if the detector and the ancilla are
uncorrelated,  it may happen that $F_{tot}> F_D$, showing that the
inclusion of an ancilla  may improve the amount of which-way information
that can be read-out from the detectors.

Since the state of the ancilla is fixed once and for all, it is possible
though to express the probabilities of the possible  outcomes resulting
from the measurement of any observable $A_{tot}$ in ${\cal H}_{tot}$, in
terms of quantities defined directly in ${\cal H}_D$. We let
$P_{\mu}\;,\;\;\mu=1,\dots,N$ the orthogonal decomposition of the identity
in ${\cal H}_{tot}$, relative to  $A_{tot}$ (we consider for simplicity an
observable with a finite number $N$ of distinct outcomes). Then,    the
probability $P_{i
\mu}$ that the
outcome $\mu$ is observed, in the state $|\chi_i>\otimes \;|\phi_{0}>$ is
given by the well known formula:
\be
P_{i\mu}= {\rm Tr}\;[P_{\mu} (\rho_i\otimes \rho_{aux})]
\ee
where $\rho_i=|\chi_i><\chi_i|$ and $\rho_{aux}=|\phi_o><\phi_0|$. If the
trace is performed in two steps, first on the ancillary Hilbert space and
then on ${\cal H}_D$, we can rewrite the above expression as
\be
P_{i\mu}= {\rm Tr}\;[A_{\mu} \;\rho_i]\;,\label{anc}
\ee
where
\be
A_{\mu}=Tr_{aux}[P_{\mu} (1\otimes
\rho_{aux})]\;,
\ee
and $Tr_{aux}$   denotes the partial trace over the ancilla Hilbert space.
The hermitean operators $A_{\mu}$ belong to ${\cal A}_D$, and it is easy to
check that they are positive definite, and that they provide a
decomposition of the identity on ${\cal H}_D$:
\be
\sum_{\mu} A_{\mu}=1
\;,\;\;\;\;{\rm on} \;\;\;{\cal H}_D
\ee
 However, in general, they
are not projection operators, neither they commute with each other. We
point out also that the number $N$ of different outcomes needs not be the
same as neither the number $n$ of detector-states, nor the dimensionality
of ${\cal H}_D$. The collection $\{A_{\mu}\}$ of operators constitutes an
example of a positive operator-valued measure (POVM) in ${\cal H}_D$. More
generally \cite{helstrom, peres}, a POVM is a map that associates to every
(Borel) subset $\Delta$ of the real line $R$, a non-negative (self-adjoint)
operator $\Pi(\Delta)$, such that:\\
\noindent
i) the empty set $\emptyset$ is mapped to zero;\\
\noindent
ii) the entire real line is mapped to the identity operator:\\
\noindent
iii) the union of any number of disjoint sets   is mapped to the sum of the
corresponding operators.\\
\noindent
The probability $P(\Delta)$ for the outcome  to be in the set $\Delta$ is
given by the following expression, generalizing equation (\ref{anc}):
\be
P(\Delta)=Tr \;[\rho \,\Pi(\Delta)]\;\;.
\ee
The axioms i), ii) and iii) listed above ensure the consistency of the
above probabilistic interpretation. POVM's thus represent a
generalization of the  projection-valued measures (PVM), usually
considered in Quantum Mechanics, and it is a theorem due to Neumark
\cite{neum}, that all POVM's on ${\cal H}_D$ can be realized by means
of an appropriate ancillary system, in the way sketched above. Since
any quantum system not interacting with the detector can play the r\^{o}le
of the ancilla, this theorem implies that every POVM can be realized
by an experimental procedure falling within the usual framework of
Quantum Mechanics. Thus, in order to determine what is the maximum
amount of which-way information that can  obtained by observing the
detector, we should maximize $\bar F$ over the set of all POVM's in
${\cal H}_D$, and not just over the set of all PVM's.\\ It is time now
to define precisely the average which-way information $\bar F$
delivered by a POVM.  For any POVM $\{A_{\mu}\;,\mu=1,\dots,N\}$ (we
shall always consider POVM with a finite number $N$ of different
outcomes, in what follows), consider the a posteriori probabilities
$Q_{i
\mu}$ for observing the $\mu-$th outcome, when the detector is in the state
$|\chi_i\!>$. According to Bayes' formula:
\be
Q_{i\mu}=\frac{\zeta_i P_{i\mu}}{q_{\mu}}\;,\label{post}
\ee
where $q_{\mu}$ is the a priori probability for the occurrence of the
outcome $\mu$:
\be
q_{\mu}=\sum_{i}\zeta_i P_{i\mu}\;.\label{apri}
\ee
In order to  measure the  amount of which-way information, that is gained
if the $\mu$-th outcome is observed, we consider the quantity
$F_{\mu}=F(\vec{Q}_{\mu})$, where $\vec{Q}_{\mu}=(Q_{1\mu}
\dots, Q_{n\mu})$ and $F$ is some function.
It is reasonable to  require from $F$ the following properties:

\noindent
(1) $F$ should be invariant under any permutation of its $n$ arguments.

\noindent
(2) $F$ should reach its absolute minimum when its $N$ arguments are all
equal to $1/N$ (which corresponds to complete lack of information on the
detector state);

\noindent
(3) $F$ should reach its absolute maximum when any of its arguments is
equal to one, while all the others are equal to zero (which on the contrary
corresponds to certain knowledge of the detector state);

\noindent
(4) $F$ should be convex, i.e. for any $\lambda \in [0,1]$ it should hold:
\footnote{$F$ is said to be strictly convex, if the equality sign in
Eq. (\ref{conc}) holds if and only if the vectors $\vec{Q}_{\mu}^{\prime}$
and $\vec{Q}_{\mu}^{\prime\prime}$ coincide.}
\be
F(\lambda \vec{Q}^{\prime}+(1-\lambda)\vec{Q}^{\prime\prime}  )
\le \lambda F( \vec{Q}^{\prime}   )+
(1-\lambda)F( \vec{Q}^{\prime\prime}   )\;.\label{conc}
\ee
The intuitive meaning of this condition is clear if we interpret
$\vec{Q}^{\prime}$ and $\vec{Q}^{\prime\prime}$ as  giving the a posteriori
probabilities of $n$ alternative hypothesis, for two  distinct tests
$A^{\prime}$ and $A^{\prime \prime}$. For any $\lambda \in [0,1]$, we can
consider the combination $A_{\lambda}$ of the tests $A^{\prime}$ and
$A^{\prime \prime}$, which consists in performing randomly either
$A^{\prime}$ or $A^{\prime \prime}$, with relative probabilities $\lambda$
and $1-\lambda$, respectively. Equation (\ref{conc}) than states that the
test $A_{\lambda}$ cannot carry  more information than the weighted sum of
the informations obtained from $A^{\prime}$ and $A^{\prime
\prime}$, separately.

The overall average information delivered by the POVM is defined as the
average $\bar F$ of the numbers $F_{\mu}$, over all possible outcomes,
weighted with the a priori probabilities $q_{\mu}$:
\be
\bar{F}:=\sum_{\mu} q_{\mu} F_{\mu}\;.\label{faver}
\ee
The optimization problem   consists in searching for the POVM which
maximizes $\bar F$. Notice that, among the unknowns, we have to
consider also the number $N$ of elements of the POVM. Of course,  the
solution depends on the choice of the function $F$, above. Over the
past years, several different  choices have been adopted. For example,
as we said in the Introduction, the authors of Refs.
\cite{{zurek,peres,davies}} consider the negative of Shannon's entropy
\cite{shannon} $H$, which corresponds to taking:
\be
F_{\mu}=-H_{\mu}:=\sum_{i} Q_{i \mu} \log Q_{i \mu} \;.\label{shann}
\ee
References \cite{englert,helstrom}  use the negative of Bayes' cost
function $C$:
\be
F_{\mu}=-\,C_{\mu}:=-\sum_{i \neq j(\mu)}   Q_{i
\mu}=Q_{j(\mu)\mu}-1\;,\label{bayes}
\ee
where, for each $\mu$, $j(\mu)$ is any index such that $Q_{j(\mu)\mu}=$
Max$\{Q_{1\mu},\dots, Q_{n\mu}\}$. Finally, more recently, D\"{u}rr \cite{durr}
considered the normalized rms spread $K$:
\be
F_{\mu}=K_{\mu}:=\left[\frac{n}{n-1}\sum_i\left(Q_{i
\mu}-\frac{1}{n}\right)^2
\right]^{1/2}\;.\label{rmssp}
\ee
When $n=2$, it is easy to check that $K_{\mu}=1-2\,C_{\mu}$, and thus the
two criteria (\ref{bayes}) and (\ref{rmssp}) are inequivalent only for more
than two beams. Notice also that, while Shannon's entropy and the rms
spread are strictly convex, the Bayes cost function is only convex.

Solving the optimization problem is a difficult task, and so far no general
solution is known. However, partial results are available. For POVM's
consisting of a finite number of elements, by using the convexity of the
function $F$, it is easy to show \cite{davies} that the optimal POVM can be
chosen to consist of rank one operators, namely:
\be
A_{\mu}= |\phi_{\mu}><\phi_{\mu}|\;,\label{rankone}
\ee
where $\|\phi_{\mu}\|\le 1$. Moreover, if  ${\cal H}_D$ is finite
dimensional and  $d$ is its dimension, it has been shown \cite{davies} that
the number $N$ of elements of the optimal POVM can be taken to satisfy:
\be
d \le N \le d^2\;.\label{ineq}
\ee

\section{Two-beam interferometers.}

In this short Section, we consider a two-beam interferometer.
For such a case, as pointed out in the previous Section, the criterion
using the Bayes cost function [Eq. (\ref{bayes})] turns out to be
equivalent to that based on the rms spreads [Eq. (\ref{rmssp})]. The
quantum detection problem, with the Bayes cost function as measure of
information, is studied at length in Ref.\cite{helstrom}. There, it is
shown that, for any number $n$ of linearly independent states $|\chi_i\!>$
and arbitrary a priori probabilities $\zeta_i$, the optimal measurement is
always a PVM. Since, in two-beam interferometers, the detector states
$|\chi_1\!>$ and $|\chi_2\!>$ must be distinct, for any path discrimination
to be possible, they are necessarily linearly independent and thus it
follows, from the quoted result, that the optimal measurement is a PVM.

\noindent
To our knowledge, there is no published proof that the optimal
measurement is a PVM, even when one uses Shannon's entropy, as a
measure of the which-way information. We have proven it, in the
special case of equally populated beams, $\zeta_i=1/2$. The rather
elaborate proof can be found in the Appendix. When the populations
$\zeta_i$ are different, we have not been able to work out an
analytical proof, but a number of numerical simulations performed for
various choices of the populations, seem to indicate that the optimal
measurement is a PVM also in this general case.

In conclusion, it appears that for two-beam interferometers, both with
Bayes's cost or with Shannon's information as measures of  which-way
information, ordinary PVM's can read out the maximum which-way information
from the detectors, and recourse to ancillas is superfluous. In fact, it
turns out that the optimal PVM is the same, for both criteria (see Eq.
(\ref{pvm}) in the Appendix).

\section{Multi-beams interferometers.}
\setcounter{equation}{0}

In this Section we  study the case of multi-beam interferometers, with
$n>2$ beams. We make the simplifying assumption that ${\cal H}_D$ is
two-dimensional. This case is actually realized in experiments using beams
of spin-half particles or photons, if the path information is stored in the
internal states of the interfering particles. A further simplifying
assumption that we make is that the beams are equally populated:
$\zeta_i=1/n$.

\noindent
${\cal H}_D$  is isomorphic to $C^2$, the set of all pairs of complex
numbers. As it is well known, rays of $C^2$ can be put in one-to-one
correspondence with unit three-vectors $\hat{n}=(n^x,n^y,n^z)$, via the
map:
\be
\frac{1+\hat n\, {\cdot} \,\vec{\sigma}}{2}|\chi>=|\chi>\;,\label{block}
\ee
where $\vec{\sigma}=(\sigma_x, \sigma_y, \sigma_z)$ is a set of Pauli
matrices. Thus, assigning $n$ pure states $|\chi_i\!>$ amounts to picking
$n$ unit vectors $\hat{n}_i$ in $R^3$.
Whether the
optimal test is a PVM or rather a POVM, now depends on the choice of the
function $F$. Below, we consider in detail the three choices for $F$, Eqs.
(\ref{shann}), (\ref{bayes}) and (\ref{rmssp}), so far considered in the
literature.

\noindent
a) $F$ is the negative of Shannon's entropy $H$ [Eq. (\ref{shann})]. For
three or more beams, it is known that the optimal test, in general, is not
a PVM but rather a POVM. For example, for three states $\hat{n}_1$,
$\hat{n}_2$ and $\hat{n}_3$ forming angles of $120^{\circ}$ with each other
and such that $\sum_{i=1}^3 {\hat n}_i=0$, it has been shown \cite{peres}
that the optimal test is provided by the following POVM with three
elements:
\be
A_{i}= \frac{1}{3} (1-\hat{n}_i {\cdot} \,\vec{\sigma})\;\label{three}
\ee

\noindent
b) $F$ is the negative of Bayes' cost function $C$ [Eq. (\ref{bayes})].
Here too, the optimal test is not a PVM, but a POVM. An example is again
provided by the set of three symmetric pure states considered under case
(a) above. It is shown in \cite{helstrom} that the optimal POVM is given
this time by the following POVM with three elements:
\be
A_{i}= \frac{1}{3} (1+\hat{n}_i {\cdot}\,\vec{ \sigma})\;\;.\label{poba}
\ee
Notice that the above POVM is not the same as [Eq. (\ref{three})], which is
an example of the fact that the solution of the optimization problem
depends on the choice of $F$.

\noindent
c) $F$ is given by the rms spread $K$ [Eq. (\ref{rmssp})]. Remarkably
enough, we can show that, for any number $n$ of equally populated  beams,
the optimal test is always a PVM. This is in sharp contrast with what
happens for the two other choices of $F$ previously considered. To prove
this claim, consider an optimal POVM, $A=\{A_{\mu};\;\mu=1,\dots N\}$. We
know, from Sec. 2, that  the operators $A_{\mu}$ must be of the form
(\ref{rankone}). Using Eq. (\ref{block}), we can write:
\be
A_{\mu}=\alpha_{\mu}(1+{\hat m}_{\mu}{\cdot} \,\vec{\sigma})\;,\label{amu}
\ee
where ${\hat m}_{\mu}$ are $N$ unit three-vectors, and $\alpha_{\mu}$ are
$N$ positive numbers. The condition for a POVM, $\sum_{\mu}A_{\mu}=1$, is
then equivalent to:
\be
\sum_{\mu}\alpha_{\mu}=1\;\;\;\;,\;\;\;\;\sum_{\mu}\alpha_{\mu}{\hat m}_{\mu}=0\;.
\label{cond}
\ee
In view of Eq. (\ref{amu}), we find:
\be
P_{i\mu}:=<\chi_i|A_{\mu}|\chi_i>=\alpha_{\mu}(1+ \hat{m}_{\mu}{\cdot}
\hat{n}_i)\;.\label{easy}
\ee
Using this equation, we compute Eq. (\ref{apri}) as:
\be
q_{\mu}=\alpha_{\mu}(1+ \hat{m}_{\mu}{\cdot} \sum_i \zeta_i
\;\hat{n}_i)\;.\label{qmuo}
\ee
In order to evaluate the average information $\bar{F}(A)$ of $A$, it is
convenient to rewrite the quantities $q_{\mu}K_{\mu}$ as
\be
q_{\mu}K_{\mu}=\left[\frac{n}{n-1}\left(-\frac{q_{\mu}^2}{n}+\sum_{i=1}^n
\zeta_i^2 P_{i \mu}^2
\right)
\right]^{1/2}\;.
\ee
Upon using Eqs. (\ref{easy}) and (\ref{qmuo}) into the above formula, we
obtain, after a little algebra:
\be
q_{\mu}K_{\mu}= \alpha_{\mu}\sqrt{\frac{n}{n-1}}
\left\{-\frac{1}{n}[1+(\hat{m}_{\mu}{\cdot}\sum_i\zeta_i \hat{n}_i)^2]+
\sum_i \zeta_i^2[1+(\hat{m}_{\mu}{\cdot} \hat{n}_i)^2]+
2\hat{m}_{\mu}{\cdot}\sum_i \zeta_i\left(\zeta_i -\frac{1}{n} \right)\hat{n}_i
\right\}^{1/2}\;.
\ee
We observe now that, for equally populated beams, $\zeta_i=1/n$, the last
sum in the above equation vanishes, and the expression for $q_{\mu}K_{\mu}$
becomes invariant under the exchange of $\hat{m}_{\mu}$ with
$-\hat{m}_{\mu}$. Consider now the POVM $B=\{B_{\mu}^+,
B_{\mu}^-;\;\mu=1,\dots,N\}$, consisting of $2N$ elements, such that:
\be
B_{\mu}^{+}  =\frac{1}{2}\;A_{\mu}\;,\;\;\;\;B_{\mu}^{-}
=\frac{1}{2}\;\alpha_{\mu}(1-{\hat
m}_{\mu}{\cdot}\,
\vec{\sigma})
\ee
Of course, $q_{\mu}^{(+)}K_{\mu}^{(+)}=q_{\mu}K_{\mu}/2$, while the
invariance of $q_{\mu}K_{\mu}$ implies
$q_{\mu}^{(-)}K_{\mu}^{(-)}=q_{\mu}^{(+)}K_{\mu}^{(+)}$. It follows
that the  average informations for $A$ and $B$ are equal to each
other, $\bar{F}(A)=\bar{F}(B)$. Now, for each value of $\mu$, the pair
of operators $B^{{\pm}}_{\mu}/\alpha_{\mu}=(1{\pm}{\hat
m}_{\mu}{\cdot}\vec{\sigma})/2$ constitutes a PVM, and thus the POVM $B$ can
be regarded as a collection of $N$ PVM's, each taken with a
non-negative weight $\alpha_{\mu}$. But then $\bar{F}(B)$, being equal
to the average of the amounts of information provided by  $N$ PVM's,
cannot be higher than the maximum information $F_D$ delivered by a
PVM. Therefore, we have proven that $F(A)=F(B)
\le F_D$, which shows that the optimum measurement can always be effected by
a means of  PVM.

We then see that, in the multibeam case, only with D\"{u}rr's measure of
information one can dispose of the ancilla, at least for equally
populated beams.

\section{Conclusions}

When, in an interference experiment, the which-way detector states are not
mutually orthogonal, one has an incomplete knowledge of the path followed
by the interfering particles. One is then faced with the problem of reading
out, in an optimum way, the information stored in the detectors. The best
measurement to be performed depends, in a crucial way, on the criterion
used to measure the information. This is a problem in quantum decision
theory, and our paper is a contribution to the task of identifying the
optimum quantum test, for which no general solution is known so far.

We have shown that for the two beams case, both by using Shannon
entropy or Bayes cost function as measures of information, the best
test to be performed is given by an ordinary projection valued
measurement in the detector's Hilbert space. Actually, it turns out
that both criteria identify the same measurement. In the multibeam
case only D\"{u}rr's normalized rms spread criterion leads to a PVM, while
the other two lead to a POVM. Notice that in the case of three
coplanar symmetric beam states one ends up with two different POVM's:
the one relative to Bayes cost [Eq. (\ref{poba})], allows every time
to pick one beam as the most probable one, while the POVM determined
by Shannon entropy, allows to exclude one of the three beams as
impossible.

We see, then, that in the multibeam case D\"{u}rr's criterion seems to be
favoured for two different reason. First of all, it allows  to derive
a quantitative complementarity relation, as the one given by Eq.
(\ref{D+V}). Second, it allows to work with ordinary quantum
mechanical measurements, and to ignore generalized POVM's, involving
an ancillary system. A possible relationship of these two features
seams worth studying. This may be related to the fact that, as has
been recently shown \cite{luis}, there are problems in extending the
mathematical definition of complementarity to a POVM.

Our results are of limited generality in two respects: first, in the
multibeam case they refer to two-state detectors, second, we always
considered equally populated beams. For what concern the latter problem, we
may add that we have gathered substantial numerical evidence that our
results may extend to arbitrarily populated beams. However we lack at the
moment an analytic proof. The former limitation seems more difficult to
overcome. Fortunately, however, the case we have treated is physically
interesting, for it includes many experimental setups in which the
"which-way" detection exploits some two-states internal degrees of freedom
of the interfering particles.

\section{Acknowledgments}

We gratefully acknowledge an interesting discussion with Prof. G. Marmo and
Prof. E.C.G. Sudarshan. The work of G.B. was partially supported by the
PRIN {\it S.Inte.Si.}. The work of R.M. was partially supported by EC
program HPRN-EC-2000-00131, and by the PRIN "{\it Teoria dei campi,
superstringhe e gravit\`{a}}".

\section{Appendix}

In this Appendix, we prove the following

\noindent {\it Theorem:} for a
two-beams interferometer with equally populated beams, when one uses the
negative of Shannon's entropy to measure the which-way information, the
optimal measurement is provided by a  PVM (precisely described in Eq.
(\ref{pvm}) below).

\noindent
More precisely,  let $|\chi_+>$ and $|\chi_->$ be the detector states, for
the two beams. We exclude the trivial case, when $|\chi_+>$ and $|\chi_->$
are proportional, because then no path-reconstruction would be possible.
Therefore, ${\cal H}_D$ is two-dimensional and we can represent vectors in
${\cal H}_D$ by unit three vectors, according to Eq. (\ref{block}). We
loose no generality if we assume that the unit vectors $\hat{n}_+$ and
$\hat{n}_-$, associated to $|\chi_+>$ and $|\chi_->$ respectively, have the
expressions:
\be\vec{n}_+=(\sin
\theta,\; 0,\;\cos
\theta)\;\;,\;\;\;\;\;\vec{n}_-=(-\sin
\theta,\;0, \;\cos \theta)\;,\label{param}
\ee
With this parametrization for the states $|\chi_+>$ and $|\chi_->$, our
theorem states that, if the which-way information is measured by the
negative of Shannon's entropy $H$, the optimal measurement is provided by
the PVM $A$ with elements:
\be
A_{+}=\frac{1}{2}(1+\sigma_x)\;\;,\;\;\;\;A_{-}=\frac{1}{2}(1-\sigma_x)\;.
\label{pvm}
\ee
Before giving the proof of this Theorem, it is useful to prove first the
following

\noindent
{\it Lemma}: consider, in $C^{\,2}$, $n$ states $|\chi_i>$, with coplanar
vectors $\hat{n}_i$, and arbitrary populations $\zeta_i$. Then, the optimal
POVM  has elements $A_{\mu}$ of the form [Eq. (\ref{amu})], with all the
vectors ${\hat m}_{\mu}$ lying in the same plane containing the vectors
$\hat{n}_i$.

\noindent
The proof of the lemma is as follows. Let $B$ be an optimal POVM. Then we
know, from the theorems quoted in Sec. 2, that its elements must have
rank-one and so are of the form given in Eq. (\ref{amu}). Moreover, they
must satisfy the POVM conditions given by Eqs. (\ref{cond}). Suppose now
that some of the vectors $\hat{m}_{\mu}^{(B)}$ do not belong to the plane
containing the vectors $\hat{n}_i$, which we assume to be the $xz$ plane.
We show below how to construct a new POVM $A
\equiv\{{A}_{\nu}\;,\;\nu=1,\dots N+p\}$, providing not less information than $B$,
and such that the vectors $\hat{m}_{\nu}^{(A)}$ all belong to the $xz$
plane. The first step in the construction of $A$ consists in  symmetrizing
$B$ with respect to the $xz$ plane. The symmetrization is done by replacing
each element $B_{\mu}$ of $B$, not lying in the $xz$ plane, by the pair
$(B_{\mu}^{\prime},B_{\mu}^{\prime\prime})$, where
$B_{\mu}^{\prime}=B_{\mu}/2$, and $B_{\mu}^{\prime\prime}$ has the same
weight as $B_{\mu}^{\prime}$, while its vector $\hat{m}_{\mu}^{(B)\prime
\prime}$ is the symmetric of $\hat{m}_{\mu}^{(B)}$ with respect to the $xz$
plane. It is easy to verify that the symmetrization preserves the
conditions for a POVM [Eqs. (\ref{cond})]. Since all the vectors
$\hat{n}_i$ belong by assumption to the $xz$ plane, we see, from Eq.
(\ref{easy}), that the probabilities $P_{i
\mu}$ actually depend only on the projections of the vectors $\hat{m}_{\mu}^{(B)}$ in the
plane $xz$. This implies, at is easy to check, that  symmetrization with
respect to the $xz$ plane does not change the information $\bar F$. We
assume therefore that $B$ has been preliminarily symmetrized in this way.
Now we show that we can replace, one after the other, each pair of
symmetric elements $(B_{\mu}^{\prime},B_{\mu}^{\prime\prime})$ by another
pair of operators, whose vectors lie in the $xz$ plane, without reducing
the information provided by the POVM. Consider for example the pair
$(B^{\prime}_{p}, B^{\prime\prime}_{p})$. We construct the unique pair of
unit vectors ${\hat u}_{p}$ and ${\hat v}_{p}$, lying the $xz$ plane, and
such that:
\be
{\hat u}_{p}+{\hat v}_{p}=2(m_{p}^{(B)x}\;\hat i+ m_{p}^{(B)z}\;\hat
k)\;,\label{split}
\ee
where $\hat i$ and $\hat j$ are the directions of the $x$ and $z$ axis,
respectively. Notice that ${\hat u}_{p}\neq{\hat v}_{p}$. Consider now the
collection of operators obtained by replacing the pair
$(B^{\prime}_{p},B^{\prime\prime}_{p})$ with the pair
$(A_{p}^{\prime},A_{p}^{\prime\prime})$ such that:
\be
A_{p}^{\prime}=\alpha_{p}^{(B)}(1+{\hat u}_{p}{\cdot}
\,\vec{\sigma})\;\;\;,\;\;\; A_{p}^{\prime\prime} =\alpha_{p}^{(B)}(1+{\hat
v}_{p}{\cdot} \,\vec{\sigma})
\;\;.
\ee
It is clear, in view of Eqs. (\ref{split}), that the new collection of
$N+p$ operators still forms a resolution of the identity, and thus
represents a POVM.  Equations (\ref{split}) also imply:
$$
P_{i p}^{(B)\prime}=P_{i p}^{(B)\prime\prime}=
\alpha_{p}(1+m^{(B)x}_{p}n^x_i+ m^{(B)z}_{p}n^z_i)=
$$
\be
=\frac{1}{2}\alpha_{p}(1+u^x_{p}n^x_i+
u^z_{p}n^z_i)+\frac{1}{2}\alpha_{p}(1+v^x_{p}n^x_i+
v^z_{p}n^z_i)=\frac{1}{2}( P_{ip}^{(A)\prime}+P_{ip}^{(A)\prime\prime})\;,
\ee
Now, define $\lambda^{\prime}_{p}:=q^{(A)\prime}_{p}/(2q_{p}^{(B)})$,
 and
$\lambda^{\prime\prime}_{p}:=q^{(A)\prime\prime}_{p}/(2q_{p}^{(B)})$, where
$q_{p}^{(B)}:= q_{p}^{\prime(B)}=q_{p}^{\prime\prime(B)}$. Since $
q_{p}^{(A)\prime}+q_{p}^{(A)\prime\prime}=2q_{p}^{(B)}$, we have
$\lambda^{\prime}_{p}+\lambda^{\prime\prime}_{p}=1$. It is easy to verify,
using Eqs. (\ref{post}) and (\ref{apri}), that:
\be
 Q_{i p}^{(B)\prime}= Q_{i p}^{(B)\prime\prime}=\lambda^{\prime}_{p}\;Q_{i
p}^{(A)\prime}+\lambda^{(A)\prime\prime}_{\mu}\;Q_{i
p}^{(A)\prime\prime}\;\;,
\ee
But then, the convexity of $F$ implies:
$$
q_{p}^{\prime(B)}F^{(B)}(\vec{Q}_p^{\prime(B)})+
q_{p}^{\prime\prime(B)}F^{(B)}(\vec{Q}_p^{\prime\prime(B)})= 2
q_{p}^{(B)}F^{(B)}(\vec{Q}_p^{\prime(B)})=
$$
$$
=2
q_{p}^{(B)}F(\lambda^{\prime}_{p}\vec{Q}_{
p}^{(A)\prime}+\lambda^{(A)\prime\prime}_{\mu}\vec{Q}_{
p}^{(A)\prime\prime})\le2 q_{p}^{(B)}[
\lambda^{\prime}_{p}F( \vec{Q}_{ p}^{(A)\prime})+
\lambda^{\prime\prime}_{p}F( \vec{Q}_{ p}^{(A)\prime\prime})]=
$$
\be
=
q_{p}^{\prime(A)}F^{(A)}(\vec{Q}_p^{\prime(A)})+
q_{p}^{\prime\prime(A)}F^{(A)}(\vec{Q}_p^{\prime\prime(A)})\;.
\ee
It follows that the new POVM is no worse than $B$. By repeating this
construction $p$ times, we can obviously eliminate from $B$ all the $p$
pairs of elements not lying in the $xz$ plane,  until we get a POVM $A$,
which provides not less information than $B$, whose elements all lie in the
$xz$ plane. This concludes the proof of the lemma.

\noindent
We can turn now to the proof of the Theorem, stated at the beginning of
this Appendix. The proof consists in showing that the PVM $A$ in [Eq.
(\ref{pvm})] provides not less information than any other POVM, $C$,
consisting of more than two elements. By virtue of the lemma just proven,
we loose no generality if we assume that the $N>2$ vectors $m_{\mu}^{(C)}$
of $C$ lie in the $xz$ plane. Our first move is to symmetrize $C$ with
respect to $z$ axis, by introducing a POVM $B$, consisting of $N$ pairs of
elements $(B_{\mu}^{\prime}, B_{\mu}^{\prime\prime})$, having equal
weights, and vectors $\hat {m}_{\mu}^{\prime}$ and $\hat
{m}_{\mu}^{\prime\prime}$ that are symmetric with respect to the $z$ axis:
\be
B^{\prime}_{\mu}=\frac{1}{2}\;C_{\mu}\;\;,\;\;\;\;B^{\prime\prime}_{\mu}=
\frac{1}{2}\;\alpha_{\mu}^{(C)}(1-{\hat m}_{\mu}^x{\sigma}_x+
{\hat m}_{\mu}^z{\sigma}_z)\;\;\;,\;\;\;
\mu=1,\dots,N\;\;.
\ee
$B$ provides as much information as $C$. Indeed, in view of Eq.
(\ref{easy}), we find
\be
P_{{\pm} \mu}^{(C)}=2\;P_{{\pm} \mu}^{(B)\prime}=2\;P_{\mp
\mu}^{(B)\prime\prime}\;\;\;,\;\;\;
\mu=1,\dots,N\;\;\;.
\ee
The invariance of $F$ with respect to permutations of its arguments, then
ensures that $\bar{F}(B)=\bar{F}(C)$. Thus, we loose no information if we
consider a POVM $B$, that is symmetric with respect to the $z$ axis. Now we
describe a procedure of reduction that, applied to a symmetric POVM like
$B$, gives rise to another symmetric POVM $\tilde B$, which contains two
elements less than $B$, but nevertheless  gives no less information than
$B$. The procedure works as follows: we pick at will two pairs of elements
of $B$, say $(B_{N}^{\prime}, B_{N}^{\prime\prime})$ and
$(B_{N-1}^{\prime}, B_{N-1}^{\prime\prime})$ and consider the unique pair
of symmetric unit vectors $\hat{u}_{{\pm}}={\pm}u^x
\;\hat i + u^z\;
\hat k$ such that:
\be
{u}^z=\frac{1}{\alpha_N^{(B)}+\alpha_{N-1}^{(B)}}  (\alpha_N^{(B)}\;
m_{N}^{(B) z} +
 \alpha_{N-1}^{(B)}\; m_{N-1}^{(B)z})\;.\label{vecu}
\ee
Consider the symmetric collection $\tilde B$, obtained from $B$ after
replacing the four elements $(B_{N}^{\prime},
B_{N}^{\prime\prime},B_{N-1}^{\prime}, B_{N-1}^{\prime\prime})$ by the pair
$(\tilde{B}_{N-1}^{\prime},\tilde{B}_{N-1}^{\prime\prime})$ such that:
\be
\tilde{B}_{N-1}^{\prime}=(\alpha_N^{(B)}+\alpha_{N-1}^{(B)})(1+
\hat{u}_+ {\cdot}\, \vec{\sigma})\;,\;\;\;\;
\tilde{B}_{N-1}^{\prime\prime}=(\alpha_N^{(B)}+\alpha_{N-1}^{(B)})(1+
\hat{u}_- {\cdot} \,\vec{\sigma})\;.
\ee
$\tilde B$ is still a POVM, as it is easy to verify. Moreover, $\tilde B$
provides not less information than $B$, as we now show. Indeed, after some
algebra, one finds:
\be
\frac{\bar{F}(\tilde B)-\bar{F}(B)}{\alpha_N^{(B)}+\alpha_{N-1}^{(B)}}=
 g(u^z)-
\frac{\alpha_N^{(B)}}{\alpha_N^{(B)}+\alpha_{N-1}^{(B)}}g(m_N^{(B)z})-
\frac{\alpha_{N-1}^{(B)}}{\alpha_N^{(B)}+\alpha_{N-1}^{(B)}}
g(m_{N-1}^{(B)z})
\;,\label{last}
\ee
where the function $g(x)$ has the expression:
$$
g(x)=(1+x \cos \theta)\log(1+x \cos \theta)+
$$
$$
-\frac{1}{2}(1+x \cos \theta+
(1-x^2)^{1/2}\sin \theta)\log \left[\frac{1}{2}(1+x \cos \theta+
(1-x^2)^{1/2}\sin \theta)\right]+
$$
\be
-\frac{1}{2}(1+x \cos \theta-
(1-x^2)^{1/2}\sin \theta)\log\left[ \frac{1}{2}(1+x \cos \theta-
(1-x^2)^{1/2}\sin \theta)\right]\;.
\ee
In view of Eq. (\ref{vecu}), the r.h.s. of Eq. (\ref{last}) is of the form
\be
g(\lambda x_1 + (1-\lambda)x_2)-\lambda \,g(x_1) -(1-\lambda)
\,g(x_2)\;,\label{form}
\ee where
$\lambda=\alpha_N^{(B)}/(\alpha_N^{(B)}+\alpha_{N-1}^{(B)})$, while
$x_1=m_N^{(B)z}$ and $x_2=m_{N-1}^{(B)z}$. It may be checked that, for all
values of $\theta$, $g(x)$ is concave, for $x\in [-1,1]$, and so the r.h.s.
of Eq. (\ref{form}) is non-negative for any value of $\lambda \in [0,1]$.
This implies that the r.h.s. of Eq. (\ref{last}) is non-negative as well,
 and so $\bar{F}(\tilde B)\ge
\bar{F}(B)$. After $N-1$ iterations of this procedure, we end up with a
symmetric POVM consisting of two pairs of elements $(B_{1}^{\prime},
B_{1}^{\prime\prime})$ and $(B_{2}^{\prime}, B_{2}^{\prime\prime})$. But
then, the conditions for a POVM, Eqs. (\ref{cond}), imply that the quantity
between the brackets on the r.h.s. of Eq. (\ref{vecu}) vanishes,  and so
Eq. (\ref{vecu}) gives $u^z=0$. This means that the last iteration gives
rise precisely to the PVM $A$ in [Eq. (\ref{pvm})]. By  putting everything
together, we have shown that $\bar{F}(C) = \bar{F}(B)\le
\bar{F}(\tilde{B})\dots \le \bar{F}(A)$, and this is the required result.

\end{document}